\documentclass[pra,reprint,amsmath,amssymb,aps,superscriptaddress]{revtex4-1}

\usepackage{amsthm}
\usepackage{algorithm}
\usepackage{algpseudocode}
\usepackage{tabularx}
\usepackage{graphicx}
\usepackage{dcolumn}
\usepackage{bm}
\usepackage{float}
\usepackage{appendix}
\usepackage{comment}
\usepackage[shortlabels]{enumitem}
\usepackage[colorlinks, linkcolor=blue,anchorcolor=blue,citecolor=blue,urlcolor=blue]{hyperref}

\usepackage[dvipsnames]{xcolor}
\usepackage{soul}

\begin{document}

\title{Efficient algorithms to solve atom reconfiguration problems.\\I.~The redistribution-reconfiguration (red-rec) algorithm}

\author{Barry Cimring}
\affiliation{Institute for Quantum Computing, University of Waterloo, Waterloo, Ontario N2L 6R2, Canada.}
\author{Remy El Sabeh}
\affiliation{Department of Computer Science, American University of Beirut, Lebanon.}
\author{Marc Bacvanski}
\affiliation{Khoury College of Computer Science, Northeastern University, United States.}
\author{Stephanie Maaz}
\affiliation{David R. Cheriton School of Computer Science, University of Waterloo, Waterloo, Ontario N2L 3G1, Canada.}
\author{Izzat El Hajj}
\affiliation{Department of Computer Science, American University of Beirut, Lebanon.}
\author{Naomi Nishimura}
\affiliation{David R. Cheriton School of Computer Science, University of Waterloo, Waterloo, Ontario N2L 3G1, Canada.}
\author{Amer E. Mouawad}
\affiliation{Department of Computer Science, American University of Beirut, Lebanon.}
\affiliation{David R. Cheriton School of Computer Science, University of Waterloo, Waterloo, Ontario N2L 3G1, Canada.}
\affiliation{University of Bremen, Bremen, Germany}
\author{Alexandre Cooper}
\email[]{alexandre.cooper@uwaterloo.ca}
\affiliation{Institute for Quantum Computing, University of Waterloo, Waterloo, Ontario N2L 6R2, Canada.}

\date{\today}

\begin{abstract}
We propose the redistribution-reconfiguration~(red-rec) algorithm to prepare large configurations of atoms using arrays of dynamic optical traps. Red-rec exploits simple heuristics and exact subroutines to solve atom reconfiguration problems on grids. It admits a fast and efficient implementation suitable for real-time operation.
We numerically quantify its performance using realistic physical parameters and operational constraints, both in the absence and presence of loss. Red-rec enables assembling large configurations of atoms with high mean success probability. Fast preparation times are achieved by harnessing parallel control operations that actuate multiple traps simultaneously. Faster preparation times are achieved by rejecting configurations of atoms containing fewer atoms than a given threshold.
However, the number of traps required to prepare a compact-centered configuration of atoms on a grid with a probability of 0.5 scales as the 3/2 power of the number of desired atoms. This finding highlights some of the challenges associated with scaling up configurations of atoms beyond tens of thousands of atoms.
\end{abstract} 

\maketitle 

\section{Introduction}
Quantum many-body systems formed by configurations of individual quantum particles, such as neutral atoms, charged ions, and molecules, offer valuable features for Quantum Information Processing (QIP)~\cite{Monroe2002, Jessen2004,Soderberg2009,Saffman2010,Kaufman2021}.
Operated as programmable quantum simulators~\cite{Feynman1982, Lloyd1996,Gross2017,Schafer2020,Browaeys2020,Monroe2021,Morgado2021,Daley2022}, these systems realize lattice spin models, which can also be mapped onto other physical models relevant for applications. 
Realizing these models provides access to their static and dynamic properties in regimes inaccessible to classical simulators.

Realizing QIP protocols often requires initializing the system in a deterministic state. For neutral atoms, this requirement implies assembling a configuration of individual atoms according to some predefined spatial geometry. Programmable arrays of optical traps address this requirement by enabling the assembly and dynamic rearrangement of individual atoms in optical traps. 



Finding sequences of displacement operations to prepare specific configurations of atoms from arbitrary ones requires solving atom reconfiguration problems~\cite{Schymik2020, Cooper2024}. These problems are hard combinatorial optimization problems, each with a potentially exponentially large number of valid solutions. These problems are even harder when considering that atoms might be lost during control operations. 

Exact and approximation algorithms that admit efficient implementations are known only for the simplest cases~\cite{Calinescu2007, Cooper2024}. These algorithms might not scale well with the problem size, preventing an implementation fast enough for real-time operation. For example, in the absence of loss, exact algorithms exist for minimizing the total distance traveled by all atoms. The Hungarian algorithm~\cite{kuhn1955hungarian,edmonds1972theoretical,Lee2017} is such an exact algorithm with a complexity of $\mathcal{O}(N_t^3)$, where $N_t$ is the number of traps.

Heuristic algorithms~\cite{Schymik2020, Sheng2021, Mamee2021, Ebadi2021, Tao2022, Tian2023} address the limitations of exact and approximation algorithms. They trade off guarantees of optimal performance for operational simplicity and computational speed. Moreover, they exploit \emph{ad hoc} principles derived from intuition and experience to define specific control subroutines. For example, parallel control operations might be exploited to speed up operational runtime~\cite{Ebadi2021,Tian2023}. 

In this paper, we introduce the redistribution-reconfiguration (red-rec) algorithm. Red-rec is a simple and fast heuristic algorithm designed for preparing two-dimensional~(2D) configurations of atoms with lattice geometries. We numerically benchmark its performance against exact and approximation algorithms, both in the presence and absence of loss. We specifically focus on the problem of preparing centered-compact configurations of atoms on grids. We show that red-rec can readily be used to prepare configurations of a few thousand atoms with high success probability at a fast rate. We also describe an approach to reduce the mean wait time between the preparation of two successful configurations. This speed-up is achieved by rejecting configurations containing fewer atoms than a given threshold. However, our evaluation exposes important challenges in scaling up configurations of atoms to larger sizes. 

These results support the main three goals of this paper:
\begin{enumerate}
    \item Introducing a hardware-agnostic formalism for reconfiguration problems to facilitate the shared development of reconfiguration algorithms. 
    \item Providing a systematic approach to quantify the performance of reconfiguration algorithms in the absence and in the presence of loss.
    \item Describing the red-rec algorithm for preparing compact-centered configurations of atoms on grids and other lattice geometries.
\end{enumerate}

The presentation of our results proceeds as follows.
We first introduce the theory of atom reconfiguration problems~(Sec.~\ref{sec:section_2}), as well as a benchmarking method to numerically evaluate the performance of reconfiguration algorithms~(Sec.~\ref{sec:section_4}).
We then describe the red-rec algorithm~(Sec.~\ref{sec:section_3}) and quantify its performance at preparing centered-compact configurations of atoms in 1D chains~(Sec.~\ref{sec:section_5}) and 2D grids~(Sec.~\ref{sec:performance2d}), both in the absence and presence of loss. We then introduce and characterize an approach to decrease the mean wait time between two successful configurations of atoms~(Sec.~\ref{sec:thresholding}). We conclude by reviewing key results and outlining opportunities to further improve the red-rec algorithm~(Sec.~\ref{sec:conclusions}). In the appendices, we review exact and approximation algorithms~(App.~\ref{sec:exact_algorithms}), benchmark their performance against the red-rec algorithm in the absence of loss~(App.~\ref{sec:exact_benchmarking}), and prove the correctness of the red-rec algorithm~(App.~\ref{app:termination_proof}). 

\section{Atom reconfiguration problems}\label{sec:section_2}

To facilitate the shared development of reconfiguration algorithms, we begin our discussion with formal definitions of fundamental concepts. These definitions aim to reduce ambiguity in nomenclature, e.g., distinguishing configurations of atoms from arrays of optical traps.

We refer to an \emph{atom} at $\vec{x}\in\mathbb{R}^3$ by $a(\vec{x},\tilde{p})$, where $\tilde{p}\in[0,1]$ is the probability of detecting the atom by performing a perfect measurement at $\vec{x}$. A \emph{perfect measurement} at $\vec{x}\in\mathbb{R}^3$, $\pi(\vec{x})$, detects an atom at $\vec{x}$ if and only if there is an atom at $\vec{x}$. If a perfect measurement detects an atom at $\vec{x}$, then the probability $\tilde{p}$ of $a(\vec{x},\tilde{p})$ is updated from a probabilistic value $\tilde{p}\in[0,1]$ to a deterministic value $p\in\{0,1\}$. An atom $a(\vec{x},p)$ with $p=0$ is said to have been lost and can be disregarded.

A \emph{configuration of atoms} $\tilde{\mathcal{C}}=\{a_j(\vec{x}_j,\tilde{p}_j)\}_{j=1}^{N_a}$ is defined as a collection of $N_a$ atoms that can each be detected by a collection of perfect measurements performed at the location of each atom, $\Pi=\{\pi_j( \vec{x}_j)\}_{j=1}^{N_a}$.
A \emph{deterministic} configuration of atoms $\mathcal{C}$ is a collection of atoms whose detection probability is unity, i.e.,~$p_j=1$ for~$1\leq j\leq N_a$.

We denote a \emph{trap} at $\vec{x}\in\mathbb{R}^3$ by $t(\vec{x},\tilde{p})$, where $\tilde{p}\in[0,1]$ is the probability that the trap $t$ contains an atom $a(\vec{x},\tilde{p})$. We set the probability $\tilde{p}$ that $t$ contains an atom to the probability of detecting $a(\vec{x},\tilde{p})$ at $\vec{x}$. 

We define a \emph{trap array} $\mathcal{A}(V,S)=\{t(\vec{x}_j,\tilde{p}_j)\}_{j=1}^{N_t}$ as a collection of $N_t=|V|$ traps. Here, $V=\{\vec{x}_j\}_{j=1}^{N_t}$ specifies the \emph{geometry} of the trap array and $S=\{\tilde{p}_j\}_{j=1}^{N_t}$ specifies the \emph{occupation state} of the trap array. 
The trap array $\mathcal{A}$ contains the configuration of atoms~$\tilde{\mathcal{C}}=\{a(x_j,\tilde{p}_j)~|~x_j\in V(\mathcal{A}),~\tilde{p}_j>0\}_{j=1}^{N_t}$ that can be mapped onto a deterministic configuration of atoms $\mathcal{C}$ by performing perfect measurements at the location of each trap of the array.

We distinguish a \emph{dynamic trap array} from a \emph{static trap array} depending on whether the location and physical properties of the traps can change over time. The set of all dynamic traps is chosen to contain the static trap array. In this way, an atom in a static trap can be displaced to any other static trap by actuating the dynamic trap array.
We refer to the \emph{target region} of the array as the set of static traps containing the target configuration of atoms. We refer to the \emph{storage region} as the set of all remaining static traps.

In the remainder of this paper, we focus our attention on trap arrays located in a restricted \emph{field of view}, which is defined as the restricted plane $[-L/2,L/2]^2\subset\mathbb{R}^2$ whose square side length is equal to $L\in\mathbb{R}$. This restricted plane replaces $\mathbb{R}^3$ in the previous definitions. 
Typical geometries $V$ in $\mathbb{R}^2$ include
\begin{enumerate}[(a)]
\item \emph{Bravais lattices} (oblique, rectangular, centered rectangular, square, or hexagonal), which are specified by their origin and generator vectors, 
\item \emph{sub-lattices}, which are specified by their origin, generator vectors of the parent lattice, and coordinate numbers (indicating which elements in the parent lattice are present), and
\item \emph{arbitrary geometries}, which are specified by a list of spatial coordinates. 
\end{enumerate}

We generally refer to a column as the primary generator vector of the static trap array along which the number of traps is the largest. When the generator vectors of the lattice are not orthogonal, then columns and rows can be understood as distinct primary and secondary generator vectors of the array. 
 
\subsection{Reconfiguration protocols}\label{sec:rec_protocols}

Atom reconfiguration problems are \emph{hidden} \emph{stochastic} problems: the presence of an atom in each trap is the realization of a random variable~(\emph{stochastic}) whose value remains unknown until a measurement is performed~(\emph{hidden}). These problems differ from \emph{deterministic} reconfiguration problems~\cite{Nishimura2018,Heuvel2013,Ito2011,Bousquet2022}, which have been studied in fundamental, graph-theoretic contexts~\cite{Fujita2015, Kato2019} and operational, application-specific contexts~\cite{Han2017, Demaine2019}. 

An atom reconfiguration problem seeks a sequence of control operations to prepare a predetermined target configuration of atoms $\mathcal{C}_T$ from an arbitrary one.
An atom reconfiguration problem is said to be \emph{solvable} if $\mathcal{C}_T$ is reachable in a finite number of control steps. Furthermore, it is \emph{efficiently solvable} if it is solvable and the number of control steps in a solution is polynomial in the size of the static trap array (the number of static traps, $N_t$). 
An \emph{optimal solution} with respect to a given cost function is a solution that minimizes the cost function, e.g., the total number of displacement operations.

\begin{figure}[t]
\includegraphics[]{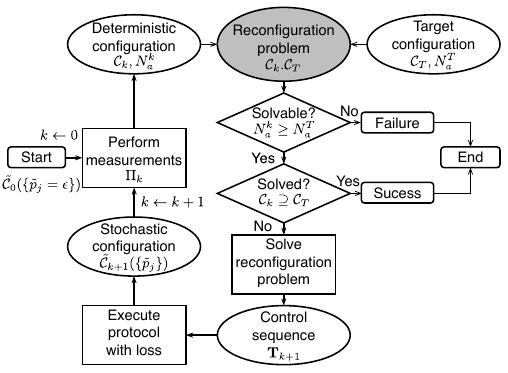}
\caption{
\label{fig:flowchart}
\textbf{Flowchart of a typical reconfiguration protocol.}
The reconfiguration protocol starts by performing measurements on a given configuration of atoms. The initial configuration of atoms  ($\tilde{\mathcal{C}}_0$) is obtained by adding an atom in each trap of the static trap array and equating the corruption of each atom ($\tilde{p}_j$) to the loading efficiency ($\epsilon$). 
The collection of measurements returns a deterministic configuration of atoms. A reconfiguration problem is formulated using the measured and desired target configuration of atoms ($\mathcal{C}_T$).
The protocol first checks if the problem is solvable. If there are less atoms than desired, the protocol declares failure and terminates. Else, the protocol checks if the problem is already solved. If the measured configuration contains the desired configuration, the declares success and ends. Else, if the problem is solvable, but not yet solved, the protocol solves the reconfiguration problem using the specified reconfiguration algorithm. The algorithm returns a control sequence ($\bm{T}_{k+1}$) to be executed on a real or simulated system. The execution of the protocol results in atom loss. Each atom in the resulting configuration of atoms ($\tilde{\mathcal{C}}_{k+1}(\{\tilde{p}\})$) is corrupted, i.e., the $j$th atom might have been lost with probability $\tilde{p}_j$. Starting with a collection of measurements, the protocol then undergoes another reconfiguration cycle until it ends in failure or success.
}
\end{figure}

The solution of an atom reconfiguration problem is a \emph{reconfiguration protocol} $\mathcal{R}=(\Pi_0, \bm{T}_1, \Pi_1, \ldots, \bm{T}_K, \Pi_K)$. This reconfiguration protocol is defined as an initial measurement $\Pi_0$ followed by a sequence of $K$ reconfiguration cycles, each cycle comprising a control sequence, $\bm{T}_k$, and a perfect measurement, $\Pi_k$. The flowchart of a typical reconfiguration protocol is presented in Fig.~\ref{fig:flowchart}.

The $k$th control sequence is a sequence of elementary control operations, $\bm{T}_k=(T_{k,1},T_{k,2},\ldots,T_{k,L_k})$, where $T_{k,l}$ is the $l$th elementary control operation of the $k$th reconfiguration cycle and $L_k$ is the number of elementary control operations required to execute the $k$th control sequence. Each elementary control operation $T_{k,l}$ is chosen from a set of six elementary control operations, $\{T_{\alpha,\nu}^{\pm,0}\}$, defined as follows:
\begin{itemize}
\item{} \emph{transfer operations}, $T_\alpha^\pm$, which include \emph{extraction operations}, $T_\alpha^+$, and \emph{implantation operations}, $T_\alpha^-$, that transfer atoms from (and to, respectively) the static trap array to (and from, respectively) the dynamic trap array, \item{} \emph{displacement operations}, $T_{\nu_{j}}^{\pm}$, that displace a dynamic trap by an elementary step along the generator vector $\vec{k}_{j}$ of the static trap array, either forward, $T_{\nu_{j}}^+$, or backward, $T_{\nu_{j}}^-$, and 
\item{} \emph{no-op operations}, $\{T_\alpha^0$, $T_\nu^0\}$, that leave some traps undisturbed while transfer and displacement operations are performed on other traps. 
\end{itemize}
Explicitly defining no-op operations facilitates accounting for the loss of idle atoms in both static and dynamic traps.

The $k$th \emph{measurement step} reveals the configuration of atoms contained in the static trap array, effectively projecting $\tilde{\mathcal{C}}_k$ into the deterministic configuration $\mathcal{C}_k$ of $N_a^k$ atoms. 
If the number of detected atoms is less than the number of desired atoms, $N_a^k<N_a^T$, then the reconfiguration problem is said to be unsolvable. In this case, the protocol is said to have failed, and the execution of the protocol is aborted.
If the number of detected atoms is greater than or equal to the number of desired atoms, $N_a^k\geq N_a^T$, or, more generally, greater than a given threshold, $N_a^k\geq N_a^{\text{thresh}}\geq N_a^T$ (see Sec.~\ref{sec:thresholding}), then the reconfiguration problem is said to be solvable. 
If the problem is solvable and if the measured configuration contains the target configuration, $\mathcal{C}_k \supseteq \mathcal{C}_T$, then the problem is said to have been solved in $K=k$ reconfiguration cycles, and the execution of the protocol is terminated. 
If the problem is solvable, but has not been solved yet, then the execution of the protocol continues until the target configuration has been reached or is no longer reachable because the number of remaining atoms is no longer sufficient. After solving the problem, the surplus atoms located outside the target region are discarded, e.g., by turning off the corresponding static traps outside the target region. 

\subsection{Control operations}~\label{subsec:corruption}

Control operations might update the state $S$ of both the static trap array, $\mathcal{A}_{s}(V_s,S_s)$, and the dynamic trap array, $\mathcal{A}_d(V_d,S_d)$, as well as the geometry $V_d$ of $\mathcal{A}_d(V_d,S_d)$.
Transfer operations $T_{\alpha}^{\pm}(\vec{x}_l)$ update the state of the traps at $\vec{x}_l$ in both the static and dynamic trap arrays. Given a static trap $t_s(\vec{x},\tilde{p})$ at $\vec{x}=\vec{x}_l$ containing an atom with probability $\tilde{p}$ and an overlaid dynamic trap $t_d(\vec{x},0)$ at $\vec{x}$ containing no atom, an extraction operation transfers an atom from the static trap to the dynamic trap. An extraction operation thus updates the static and dynamic traps from $[t_s(\vec{x},\tilde{p}), t_d(\vec{x},0)]$ to $[t_s(\vec{x},0),t_d(\vec{x},\tilde{p}')]$, where $\tilde{p}'=\tilde{p}~p_\alpha\exp(-t_\alpha/t_{trap})$ is the updated state probability (see next paragraph for relevant definitions). 

Similarly, an implantation operation transfers an atom from a dynamic trap to a static trap, updating the static and dynamic traps from $[t_s(\vec{x},0), t_d(\vec{x},\tilde{p})]$ to 
$[t_s(\vec{x},\tilde{p}'), t_d(\vec{x},0)]$.
A transfer operation acting on a pair of empty traps returns empty traps, as no atom can be spontaneously created. In contrast, a transfer operation acting on a pair of occupied traps returns a pair of empty traps due to collisional loss.
These definitions assume that the static and dynamic traps are overlaid; otherwise, implanting an atom in free space leads to its loss. 

In the same vein, elementary displacement operations $T_{\nu_{j}}^{\pm}$ acting on the dynamic trap $t_d(\vec{x},\tilde{p})$ at $\vec{x}$ will displace the trap and the atom it contains from $\vec{x}$ to $\vec{x}\pm\vec{k}_{j}$, where $\vec{k}_{j}$ is the generator vector of the static trap array, e.g., $\vec{k}_{j}\in\{\vec{k}_x,\vec{k}_y\}$ with $\vec{k}_x\perp \vec{k}_y$ on the grid. 

Control operations acting on different individual traps may be executed in parallel by batching them into a sequence of control operations acting on multiple traps simultaneously. The batching rules used to batch control operations are defined by operational constraints, e.g., the maximum number of dynamic traps, and the set of allowed operations, e.g., whether non-contiguous blocks of atoms can be simultaneously displaced without affecting other trapped atoms.

In the presence of loss, each elementary control operation acting on a trap is imperfect, possibly inducing the loss of the atom it contains. We define the \emph{survival probability} for transfer and displacement operations as $p_\alpha$ and $p_\nu$, respectively. To account for the finite trapping lifetime, we multiply the survival probability by the exponential decay factor, $\exp(-t/t_{trap})$, where $t\in\{t_\alpha, t_\nu\}$ is chosen as the time duration of each operation and $t_{trap}$ is the trapping lifetime. We thus refer to the \emph{corruption} of an atom as the cumulative probability of it being lost given its control history. More formally, $\tilde{p}={p_\alpha}^{N_\alpha} {p_\nu}^{N_\nu}\exp(-t_{tot}/t_{trap})$, where $N_\alpha$, $N_\nu$ are the numbers of transfer and displacement operations performed on the atoms, and $t_{tot}=N_\alpha^\Sigma t_\alpha + N_\nu^\Sigma t_\nu$ is the total duration of the reconfiguration cycle computed from the total number $N_{\alpha,\nu}^{\Sigma}$ of batched control operations. Our calculation ensures that we do not double count operations performed in parallel. 

We choose $\tilde{p}$ to denote both the detection probability and the corruption, as the two quantities are equivalent for perfect measurements. The corruption thus corresponds to the probability of identifying an atom as lost upon performing a perfect measurement. These definitions imply a definition for the survival probability for no-op operations performed on certain atoms while others experience transfer or displacement operations, $p_0^{\alpha,\nu}=\exp(-t_{\alpha,\nu}/t_{trap})$. For simplicity, we assume that the no-op survival probability is the same for both static and dynamic traps. We also assume that control errors are limited to loss. That is, we ignore the probability of an atom being left behind in its original trap during a transfer operation. Such an error can, however, be modeled as a decrease in survival probability. 

Finally, to simplify their experimental implementation, we restrict displacements to elementary displacement operations executed in constant time along the generator vectors of the static trap array. Furthermore, to 
ensure compatibility with acousto-optic deflectors generating dynamic trap arrays, we restrict parallel control operations to chains of atoms. These chains correspond to horizontal sub-rows or vertical sub-columns in the grid of the static trap array. More elaborate strategies for parallel control operations might achieve greater performance. These strategies include simultaneously displacing contiguous or non-continguous blocks of atoms or displacing atoms along arbitrary trajectories at variable speed.

\section{Methods for quantifying performance}\label{sec:section_4}
We now describe our approach to quantify the performance of reconfiguration algorithms. We categorize the performance into three distinct types:
\begin{itemize}
    \item \textit{Operational performance} measures how fast and with which success probability a target configuration of atoms can be prepared.
    \item \textit{Runtime performance} measures how fast a specific implementation of the algorithm can be executed on a processor.
    \item \textit{Algorithmic performance} measures the complexity of the algorithm and its scaling with relevant parameters.
\end{itemize}
In the following evaluation, we specifically focus on the operational performance. Recent work, which will be reported elsewhere, indicates that red-rec admits a fast runtime implementation. 

\subsection{Operational performance in the absence of loss: Baseline success probability}
In the absence of loss, we quantify operational performance by the total number of transfer and displacement operations. We further quantify operational performance in terms of the \emph{baseline success probability}. The baseline success probability is defined as the probability that the initial configuration of atoms contains at least as many atoms as needed to prepare the target configuration of atoms, $p_0=P(N_a^0\geq N_a^T)$. The number of atoms in the initial configuration ($N_a^0$) follows a binomial distribution, $N_a^0\sim\text{Bino}(N_{t},\epsilon)$. The binomial distribution is determined by the size of the static trap array, $N_{t}$, and the mean loading efficiency, $\epsilon$. 

The baseline success probability is thus given by the complementary cumulative distribution function of the binomial distribution, $p_0=\text{BinoCCDF}(N_{t},\epsilon;N_a^T)$, which has a known analytical expression. 
The baseline success probability provides an exact upper bound on the mean success probability in the presence of loss. It also provides an estimate of the minimum number of static traps required to prepare a given target configuration of atoms with a given success probability. 

Quantifying performance in the absence of loss decouples the analysis from the realization of random processes and the choice of specific experimental parameters like the duration of control operations. It also provides the means to establish an upper bound on the operational performance in the presence of loss. 

Furthermore, quantifying performance in the absence of loss enables a quantitative comparison against exact and approximation algorithms (see App.~\ref{sec:exact_algorithms}). Here, we benchmark the operational performance of red-rec against two algorithms: an exact assignment algorithm and a 3-approximation algorithm (3-approx). The exact assignment algorithm exactly minimizes the total number of displacement operations. The 3-approx algorithm seeks to minimize the total number of transfer operations; it provides a solution where the total number of transfer operations is at most three times larger than the optimal value~(see App.~\ref{sec:exact_benchmarking}). 

\subsection{Operational performance in the presence of loss: Mean success probability}
In the presence of loss, we quantify the operational performance in terms of the \emph{mean success probability}, $\bar{p}$. The mean success probability cannot exceed the baseline success probability, i.e., $\bar{p}\leq p_0$. To calculate the mean success probability, we average the probability of successfully preparing a target configuration of atoms over an ensemble of randomly-sampled initial configurations of atoms and randomly-sampled realizations of loss processes. 

The mean success probability finds practical applications in remote-access scenarios for which the user is different from the operator. The user first specifies the desired configuration of atoms. The operator then confirms whether this configuration of atoms is achievable. If achievable, the operator informs the user with which probability the configuration of atoms will be prepared.  The mean success probability depends on the number of reconfiguration protocols executed and the mean wait time between two successful protocols; it thus correlates with total execution cost. 

The performance of different shared-access platforms can thus be compared by computing the mean success probability for a given target configuration or the largest configuration achievable at a fixed mean success probability. Reporting the largest configuration that can be prepared without indicating the mean number of trials required to prepare that configuration is otherwise misleading, as obtaining a configuration of that size might rely on very unlikely sampling realizations.

\subsection{Operational performance benchmarking module}\label{sec:operational_performance}
We numerically quantify the operational performance of reconfiguration algorithms using a custom-built benchmarking module that exploits the Monte Carlo sampling method implemented in Python.

The module takes as input a reconfiguration problem, a set of physical parameters, and a set of sampling parameters. The reconfiguration problem defines the static trap array, including its geometry and dimensions, and the target atom configuration. The physical parameters include the loading efficiency and the survival probability for displacement and transfer operations. The sampling parameters include the number of simulation repetitions, the number of atoms in the initial configuration (if fixed), and the name of the algorithm to be used.

First, the module constructs a configuration of $N_a^0$ atoms. If $N_a^0$ is not fixed, the module samples $N_a^0$ from the binomial distribution defined by the loading efficiency, $\epsilon$, and total number of traps, $N_t$. Given $N_a^0$, the module then randomly assigns an atom to each of $N_a^0$ randomly-chosen traps. It does so, for example, by randomly selecting $N_a^0$ integers in a list of $N_t$ trap indices.

Second, given the initial and target configurations, the module solves the reconfiguration problem using the specified algorithm. The solution is a sequence of control operations. The module performs these operations on a simulated system to compute the numbers of transfer, displacement, and idle operations for each atom. Given the history of control operations for each atom and the survival probability for each control operation, the module computes the corruption of each atom.

Third, given the corruption of each atom, the module generates a configuration of atoms that would be obtained following a perfect measurement. This configuration is computed by sampling a number $\mu$ from the uniform distribution on $[0,1]$ for each atom. If $\mu>\tilde{p}$, the atom is lost, i.e., the state of the trap containing the atom is set to 0. If $\mu\leq\tilde{p}$, the atom is kept in the trap and the state of the trap containing the atom is set to 1.

These configuration sampling and problem solving steps continue until the problem has been solved or is no longer solvable. Each data point is typically averaged over 1,000 initial configurations.

\subsection{Typical experimental parameters and limitations}\label{sec:parameters}

To quantify the expected performance under realistic experimental conditions, we numerically compute the mean success probability using realistic physical parameters.

We choose the duration of elementary transfer and displacement operations to be $t_\alpha=15~\mu\text{s}$ and $t_\nu=67~\mu\text{s}$, respectively, and the survival probability to be $p_\nu=p_\alpha=0.985$~\cite{Ebadi2021}. The survival probability typically decreases with the displacement distance; however, this decrease can typically be mitigated by increasing the displacement time~\cite{Bluvstein2021}. We conservatively choose the trapping lifetime to be $t_{trap}=60~\text{s}$, although it could be further enhanced up to at least $6{,}000~\text{s}$ using cryogenics~\cite{Schymik2021}.

We further choose the mean loading efficiency to be $\epsilon=0.60$, which is typical for rubidium-87 atoms loaded in the collisional blockade regime~\cite{Schlosser2001}. An efficiency $\epsilon\geq0.60$ has also been achieved using enhanced-loading techniques~\cite{Grunzweig2010, Carpentier2013, Lester2015, Brown2019, Aliyu2021, Angonga2022, Jenkins2022}. To quantify performance under enhanced loading, we choose an increased loading efficiency of $\epsilon=0.90$. We assume that the loading efficiency is the same for all traps in the static trap array, as can be achieved using closed-loop optimization routines~\cite{Schymik2022}.

To quantify the size of the largest configuration of atoms that could be realistically prepared given typical experimental limitations, we restrict the accessible range of trap parameters.

An important bottleneck in scaling up atom configurations beyond a few thousand atoms is the limit on the number of optical traps. Limits may be due to restrictions in optical power, diffraction efficiency of active diffractive optical devices like spatial light modulators and acousto-optic deflectors, as well as transmission efficiency of optical elements. As a realistic, near-term operational limit on the number of optical traps, we choose the maximum size of the static trap array to be $N_t^{max}=2{,}048$ traps and set the stretch value to twice that amount, i.e., $N_t^{max}=4{,}096$. 

A trap array with such a number of traps could be realized with approximately $5~\text{W}$ in laser power at the source given $1~\text{mW}$ per optical trap in the focal plane of the optical system and a total power delivery efficiency of $0.4$ from the laser source. This power value is typical for a Ti:Sapphire laser operating at $813~\text{nm}$ that is pumped with an $18~\text{W}$ pump laser at $532~\text{nm}$. Further increase in the number of traps might be achieved by exploiting advances in laser technology. Examples include pumping Ti:Sapphire lasers at a higher power, using multiple laser sources, either coherently combined or not (e.g., if operating at a magic trapping wavelength is not crucial), or using high-power fiber lasers and fiber amplifiers.

Another bottleneck is restrictions in the field of view (FOV) of the high-resolution microscopy system used to produce optical traps and image individual atoms. We upper bound the width and height of the trap arrays by $N_{t}^{y, max}=N_{t}^{x, max}=100$ traps. Such bounds result, for example, from a grid with an inter-trap spacing of $3.0~\mu\text{m}$ in an optical system with a FOV of $300~\mu\text{m}$. An important challenge in increasing the FOV of the optical system is avoiding aberration. Aberration distorts optical traps and their loading efficiency~\cite{Schymik2022}. Significant advances have been made towards solving this problem. A research group in molecular biology has recently reported on the design of a ``mesolens’’ for operation in the wavelength range of 400 nm to 750 nm with a FOV of 6 mm, a lateral resolution of $0.6~\mu\text{m}$, and a numerical aperture of 0.5~\cite{McConnell2016}.

\section{The red-rec algorithm}\label{sec:section_3}

\begin{figure*}[t!]
\includegraphics{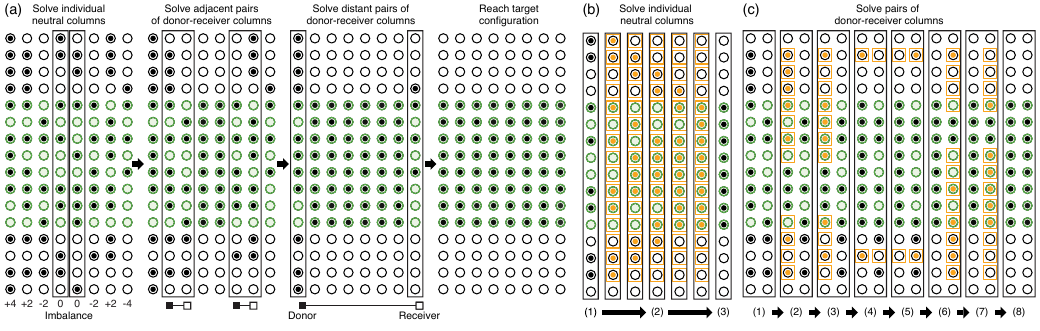}
\caption{\label{fig:reconfiguration_protocol}
\textbf{Single realization of the red-rec algorithm in the absence of loss.}
(a)~Sequence of control operations to prepare a target configuration of $N_a^T=8\times8$ atoms (green shaded disks) in a static trap array of $N_t=8\text{ (width of the trap array) }\times16\text{ (height of the trap array)}=128$ traps (circles) from an arbitrary configuration of $N_a^0=64$ atoms (black dots).
After computing the imbalance of each column as the difference between the number of detected and desired atoms, the algorithm sequentially reconfigures individual neutral columns, neighboring pairs of donor-receiver columns, and then distant pairs of donor-receiver columns.
(b)~An individual neutral column is solved by extracting atoms from the static trap array into the dynamic trap array (orange squares), displacing all atoms to their desired target locations while avoiding collisions, and implanting the atoms back into the static trap array. 
(c)~A pair of donor-receiver columns is solved using a sequence of control operations that ((1)-(3))~reconfigures the donor column while displacing atoms to be exchanged to open rows, ((4)-(5))~redistributes surplus atoms from the donor column to the receiver column in a single parallel displacement step, and ((6)-(8))~reconfigures the receiver column.
}
\end{figure*}

In this section, we describe the various steps of the red-rec algorithm (see Alg.~\ref{alg:redrec} and Fig.~\ref{fig:reconfiguration_protocol}). 

\begin{algorithm}[H]
\caption{-- The red-rec algorithm}\label{alg:redrec}
\begin{algorithmic}
\Require A trap array $\mathcal{A}(V,S)$ with a grid geometry, the number of rows $r$, the number of columns $c$, the number of measured atoms $N_a$, the number of desired atoms $N_a^T$, and
a user-defined threshold value $N_a^\text{thresh}$, 
\If{$N_a - N_a^T < N_a^\text{thresh}$}
\State \Return ``failure''
\EndIf
\For{$j \gets 1 \text{ to } c$}
\State compute the atom imbalance for column $j$ in $\mathcal{A}$
\If{column $j$ in $\mathcal{A}$ is a neutral column}
\State solve column $j$ in $\mathcal{A}$ using the exact 1D algorithm
\EndIf
\EndFor
\While{there exists a column with negative imbalance in $\mathcal{A}$}
\State select the closest pair of donor-receiver columns in $\mathcal{A}$
\\
\hskip2.5em (tie-breaking rules are applied whenever necessary) 
\State solve the selected pair in $\mathcal{A}$ following 6 steps:
\State \hskip1.3em 1. donor extraction
\State \hskip1.3em 2. donor displacement
\State \hskip1.3em 3. donor implantation
\State \hskip1.3em 4. donor-column redistribution
\State \hskip1.3em 5. receiver extraction
\State \hskip1.3em 6. receiver displacement and implantation
\EndWhile
\For{$j \gets 1 \text{ to } c$}
\If{column $j$ in $\mathcal{A}$ is not solved}
\State solve column $j$ in $\mathcal{A}$ using the exact 1D algorithm
\EndIf
\EndFor
 \end{algorithmic}
\end{algorithm}

Before solving the problem, red-rec checks whether the problem is solvable.
Given a measured configuration of $N_a$ atoms, where we drop the cycle index $k$ to improve clarity, red-rec compares the number of measured atoms ($N_a$) to the number of desired atoms ($N_a^T$). The problem is solvable if $N_a - N_a^T \geq 0$. Additionally, red-rec checks if the atom surplus exceeds a user-defined threshold value ($N_a^\text{thresh}$), i.e., $N_a - N_a^T \geq N_a^\text{thresh}$. If the problem is unsolvable or the threshold is not met, red-rec aborts and declares failure.

If the problem is solvable and the threshold is met, red-rec proceeds to compute the atom imbalance for each column. The imbalance of the $j$th column, denoted as $\Delta N_a^j$, is calculated as $N_a^j - N_a^{T,j}$, where $N_a^j$ and $N_a^{T,j}$ represent the numbers of atoms in the $j$th column of the measured and desired configurations, respectively.  Here, we use the shorthand notation $N_a^j$ to represent $N_a^{k,j}$, where $N_a^{k,j}$ denotes the number of atoms in the $j$th column during the $k$th reconfiguration cycle. Based on the imbalance, each column is labeled as a 
\begin{enumerate}[label=(\alph*)]
    \item \emph{donor}, \text{if } $\Delta N_a^j>0$,
    \item \emph{neutral}, \text{if } $\Delta N_a^j=0$, or
    \item \emph{receiver}, \text{if } $\Delta N_a^j<0$.
\end{enumerate}


After labeling the columns, red-rec breaks down the 2D reconfiguration problem into simpler reconfiguration problems involving individual columns and pairs of columns.

First, red-rec solves each neutral column using an exact reconfiguration algorithm on chains, referred to as the ``exact 1D algorithm'' (see App.~\ref{sec:exact_algorithms}), which runs in time linear in the number of traps. When applied to an individual column, the algorithm minimizes the total number of displacement operations and ensures that each atom is extracted and implanted at most once. However, in the presence of loss, it does not guarantee the overall minimization of displacement and transfer operations across multiple reconfiguration cycles.

Solving neutral columns facilitates the redistribution of atoms among distant pairs of donor and receiver columns. Doing so creates \emph{distribution rows} in which no obstructing atoms are present between the donor and receiver columns.

Second, red-rec solves pairs of donor-receiver columns. It begins with adjacent pairs where the donor column can fully satisfy the receiver column, i.e., $\Delta N_{\text{donor}} + \Delta N_{\text{receiver}} \geq 0$. Next, it solves adjacent pairs of columns. Finally, it solves distant pairs, starting with the pairs that can exchange the highest number of atoms in a single redistribution sequence.


Pairs of donor-receiver columns are solved using one or more redistribution sequences (Fig.~\ref{fig:reconfiguration_protocol}c). To realize a redistribution sequence, the atoms in the donor column are categorized into three groups:
\begin{enumerate}
    \item \emph{Reconfigured atoms} populate the target region of the donor column. They are selected in a way that minimizes the total number of displacement operations. The identification of reconfigured atoms involves utilizing the exact 1D algorithm (see App.~\ref{sec:exact_algorithms}).
    
    \item \emph{Redistributed atoms} populate the target region of the receiver column. They are chosen from the storage region of the donor column using a heuristic. The heuristic selects atoms starting from those located farthest away from the target region. These atoms are assigned to the traps in the receiver column that are situated on the nearest distribution row. The selection of redistributed atoms aims to have approximately equal numbers of atoms above and below the target region.

    \item \emph{Idle atoms} remain undisturbed in the donor column and do not undergo any displacement.
\end{enumerate}

The redistribution sequence then involves executing the following six steps:
\begin{enumerate}
    \item \emph{Donor extraction}: The redistributed and reconfigured atoms are extracted from the donor column~(Fig.~\ref{fig:reconfiguration_protocol}c(2)).
    \item \emph{Donor displacement}: The reconfigured atoms are displaced to traps in the target region of the donor column. The redistributed atoms are displaced to traps located on distribution rows in the storage region of the donor column~(Fig.~\ref{fig:reconfiguration_protocol}c(3)).
    \item \emph{Donor implantation}:
    The reconfigured atoms in the target region are implanted into the static traps (Fig.~\ref{fig:reconfiguration_protocol}c(4)).
    Meanwhile, the redistributed atoms remain extracted in the dynamic traps.
    \item \emph{Donor-receiver redistribution}: The extracted redistributed atoms are simultaneously displaced to the receiver column~(Fig.~\ref{fig:reconfiguration_protocol}c(5)). 
    \item \emph{Receiver extraction}: The atoms in the receiver column are extracted (Fig.~\ref{fig:reconfiguration_protocol}c(6)). 
    \item \emph{Receiver displacement and implantation}: The receiver column is solved using the exact 1D algorithm (Fig.~\ref{fig:reconfiguration_protocol}c(7)).
\end{enumerate}
This approach ensures that the redistributed atoms are only extracted and implanted once, which helps minimize the overall number of transfer operations.

If not all redistributed atoms can be redistributed from the donor column to the receiver column in a single redistribution sequence, then the redistribution sequence is repeated until the donor column has donated all its surplus atoms. Such a situation could occur, for example, if the number of distribution rows is less than the number of atoms to be distributed. If the receiver column has not been fully solved, then it participates again in the donor-column pairing. 

Pairs of donor-receiver columns are thus solved using redistribution sequences until all columns achieve non-negative imbalances. In App.~\ref{app:termination_proof}, we show that this step requires solving a finite number of donor-receiver column pairs, which implies the correctness of the red-rec algorithm.

Finally, red-rec solves all columns that have not yet been solved. The termination of the second step guarantees that all such columns have positive imbalance, meaning that they can be individually solved.

\section{Performance for 1D chains}\label{sec:section_5}

\begin{figure}[t]
\includegraphics[]{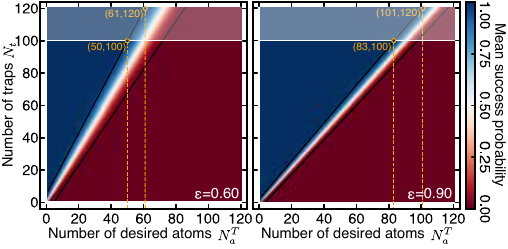}
\caption{
\label{fig:baseline_success_probability_1d}
\textbf{Baseline success probability for 1D chains.}
Probability of successfully preparing a centered-compact configuration of $N_a^T$ atoms in a 1D chain of $N_{t}$ traps for a loading efficiency of $\epsilon=0.60$ (left) and $\epsilon=0.90$ (right) in the absence of loss. A transition region separates regions of near-certain failure ($\bar{p}\leq 0.02$, red) and near-certain success ($\bar{p}\geq 0.98$, blue). The maximum number of traps is set to $N_t^{max}=100$ (shaded white region).
The transition midline at $p_0=0.5$ is realized for $N_t=N_a^T/\epsilon$ traps, showing that, in the absence of loss, the number of needed traps scales linearly with the number of desired atoms. 
}
\end{figure}

The red-rec algorithm solves atom reconfiguration problems on grids by reducing them to a sequence of reconfiguration problems on 1D chains. These problems on chains can be efficiently solved using the exact 1D algorithm. The performance of the red-rec algorithm is thus intrinsically tied to the performance of the exact 1D algorithm.

We thus start our analysis by numerically evaluating the performance of the exact 1D algorithm~(see Sec.~\ref{sec:operational_performance} for its detailed description). We perform the analysis both in the absence and presence of loss. For this analysis, we compute the baseline success probability for preparing a compact chain of $N_a^T$ atoms at the center of a chain of $N_{t}$ static traps for various values of $N_a^T$, $N_t$, and $\epsilon\in\{0.6,0.9\}$.

In the absence of loss, the baseline success probability surfaces (Fig.~\ref{fig:baseline_success_probability_1d}) exhibit a sharp transition between a region of near-certain failure~($p_0\leq0.02$) and a region of near-certain success~($p_0\geq0.98$). We note that the values delineating the regions have been chosen arbitrarily without implications for the key results of this paper.

The transition region is centered around the iso-probability line $p_0=0.5$ realized for $N_t^0=\eta_0 N_a^T=N_a^T/\epsilon$, i.e., for a $\emph{baseline overhead factor}$ $\eta_0=1/\epsilon$.
The largest centered-compact configuration of atoms that can be prepared in chains of $100~(120)$ static traps with near-certain success thus contains $N_a^T=50~(61)$ atoms for $\epsilon=0.60$ and $83~(101)$ atoms for $\epsilon=0.90$.

\begin{figure}[t]
\includegraphics[]{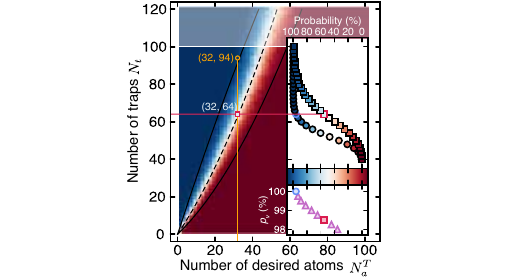}
\caption{
\label{fig:1d_loss_a}
\textbf{Mean success probability for 1D chains.}
Mean probability of successfully preparing a centered-compact configuration of $N_a^T$ atoms in a 1D chain of $N_t$ traps for a loading efficiency of $\epsilon=0.6$ in the presence of loss. The number of traps at the transition midline (dashed line) scales quadratically with the number of desired atoms, $N_t\sim \mathcal{O}((N_a^T)^2)$. Inset: Mean success probability of preparing a chain of $N_a^T=32$ atoms in the absence (disks) and presence (squares) of loss for varying success probability during control operations (bottom). At least $N_t=64$ traps are required to prepare a compact chain of $N_a^T=32$ atoms with $\bar{p}\geq0.5$.
}
\end{figure}

\begin{figure}[t]
\includegraphics[]{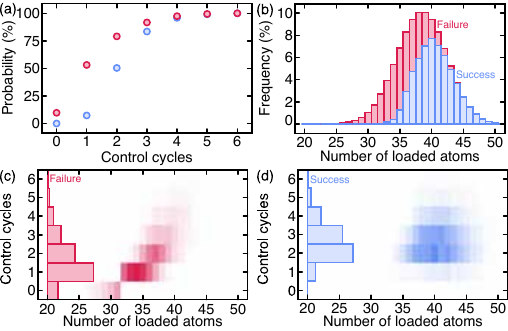}
\caption{
\label{fig:1d_loss_b}
\textbf{Control cycles for 1D chains.}
(a)~Cumulative distribution function of the number of control cycles performed during unsuccessful (red) and successful (blue) protocols for preparing a configuration of $N_a^T=32$ atoms in a chain of $N_t=64$ traps. Protocols fail faster than they succeed: half of unsuccessful protocols have failed after one cycle, whereas half of the successful protocols require at least two cycles to succeed. 
(b)~Stacked distributions of the number of atoms loaded in the static trap array. Successful protocols start with more atoms on average.
(c-d)~Distribution of the number of control cycles and the number of atoms loaded in the static array. Integrating over control cycles returns the stacked distributions of (b). Unsuccessful protocols fail in a few control cycles because the number of loaded atoms is not enough to compensate for loss.
}
\end{figure}

In the presence of loss, the mean success probability surface~(Fig.~\ref{fig:1d_loss_a}) exhibits the same two regions of near-certain success and near-certain failure as in the lossless case. There is a sharp transition region whose boundary is determined by the loading efficiency and loss parameters. 
In contrast to the lossless case, however, the transition curve is non-linear. The non-linearity indicates that an increasingly large overhead factor is required to prepare increasingly large chains of atoms with near-certain success.
We find that the overhead ratio for $\bar{p}\approx0.50$ is well-approximated by $\eta(N_a^T)\approx\eta_0+\eta_1 N_a^T$, where $\eta_0=1.50(3), \eta_1=0.014(3)$. This statement implies that $N_{t}=\eta_0 N_a^T+\eta_1 (N_a^T)^2\sim\mathcal{O}((N_a^T)^2)$ traps are required to successfully prepare a chain of $N_a^T$ atoms.
The largest centered-compact configuration of atoms that can be prepared with near-certain success in a chain of $100~(120)$ static traps thus contains $N_a^T=34~(42)$ atoms in the presence of loss, as opposed to $N_a^T=50~(61)$ atoms in the absence of loss.

Because of its relevance for preparing 2D configurations of $32\times32=1{,}024$ atoms, we further analyze the problem of preparing a chain of $N_a^T=32$ atoms in an array of $N_t=64$ traps. 
The mean success probability is equal to $\bar{p}\approx0.5$ and increases to unity as the survival probability during control operations increases to unity~(inset of Fig.~\ref{fig:1d_loss_a}). 
Focusing on successful and unsuccessful reconfiguration protocols~(Fig.~\ref{fig:1d_loss_b}), we observe that the mean number of control cycles is smaller for unsuccessful protocols than for successful protocols. Consequently, protocols are quicker to fail than to succeed:
half of unsuccessful protocols have failed after one cycle,
whereas half of the successful protocols require at least two
cycles to succeed~(Fig.~\ref{fig:1d_loss_b}a).

We also observe that successful protocols have more atoms on average in their initial configurations than unsuccessful protocols~(Fig.~\ref{fig:1d_loss_b}b). The surplus atoms are used in late reconfiguration cycles to replace atoms lost in early reconfiguration cycles. This information is also contained in the probability distribution functions for the numbers of control cycles and the numbers of atoms in the initial configuration for successful and unsuccessful protocols~(Fig.~\ref{fig:1d_loss_b}c-d). 

These observations indicate that rejecting configurations containing less than a certain number of loaded atoms might prevent the execution of protocols unlikely to succeed~(see Sec.~\ref{sec:thresholding}). For example, the probability that a control protocol is successful given that $N_a^0$ atoms have been loaded is greater than 0.5 for $N_a^0\ge37$; configurations containing fewer atoms could be rejected~(see Sec.~\ref{sec:thresholding}).

\section{Performance for 2D grids} \label{sec:performance2d}
We now quantify the performance of the red-rec algorithm at solving reconfiguration problems on 2D grids. We specifically focus on the problem of preparing a configuration of $N_a^T=\sqrt{N_a^T}\times\sqrt{N_a^T}$ atoms at the center of a square lattice of $N_t=N_{t}^x\times N_{t}^y$ static traps. Here, we choose the width of the trap array to be equal to the width of the target configuration, i.e., $N_{t}^x=\sqrt{N_{a}^T}$. We further choose the height of the trap array to contain a larger number of traps, $N_t^y=\eta\sqrt{N_{a}^T}$, where $\eta \geq 1$. Figure~\ref{fig:reconfiguration_protocol} provides an example of such a grid with rectangular dimensions.

\begin{figure}[t]
\includegraphics[]{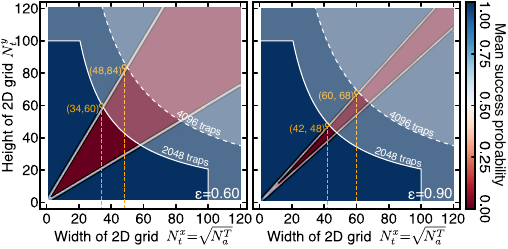}
\caption{
\label{fig:baseline_success_probability_2d}
\textbf{Baseline success probability for 2D grids.}
Probability of successfully preparing a centered-compact configuration of $N_a^T=\sqrt{N_a^T}\times\sqrt{N_a^T}$ atoms in a square-lattice array of $N_{t}^x\times N_{t}^y$ traps where $N_{t}^x=\sqrt{N_a^T}$ in the absence of loss. Increasing the loading efficiency from $\epsilon=0.60$ (left) to $\epsilon=0.90$ (right) narrows down the failure region (red).
The largest achievable configuration is ultimately restricted by the maximum number of optical traps along each direction ($N_t^{x,y}\leq100$) and in total ($N_t\leq2{,}048~ (4{,}096)$). 
As for 1D chains, the transition midline at $p_0=0.5$ is realized for $N_t=N_a^T/\epsilon$ traps ($N_t^y=\sqrt{N_a^T}/\epsilon$). In the absence of loss, the number of needed traps scales linearly with the number of desired atoms.
}
\end{figure}

In the absence of loss, the baseline success probability surfaces exhibit a region of near-certain failure $(\bar{p}\leq 0.02)$ and a region of near-certain success $(\bar{p} \geq 0.98)$~(Fig.~\ref{fig:baseline_success_probability_2d}). The transition region is much sharper than in the 1D case~(Fig.~\ref{fig:baseline_success_probability_1d}). In square-lattice trap arrays containing at most $N_{t}=2{,}048~(4{,}096)$ optical traps, the largest centered-compact configuration of atoms that can be prepared with near-certain success contains $N_a^T=34^2=1{,}156~(42^2=1{,}756)$ atoms~(Fig.~\ref{fig:baseline_success_probability_2d}a). Increasing the loading efficiency to $\epsilon=0.90$ increases the configuration size to $N_a^T=48^2=2{,}304~(60^2=3{,}600)$ atoms~(Fig.~\ref{fig:baseline_success_probability_2d}b). Assuming height limits of $100~(120)$ traps, the largest configuration of atoms that can be prepared with near-certain success contains $N_a^T=58^2=3{,}364~(70^2=4{,}900)$ atoms for $\epsilon=0.60$ and $N_a^T=88^2=7{,}744~(106^2=11{,}236)$ atoms for $\epsilon=0.90$. Given a loading efficiency of $\epsilon=0.60$, preparing a configuration of $100\times100=10{,}000$ atoms with near-certain success would thus require at least $100\times330=33{,}000$ static traps, a formidable operational challenge.

\begin{figure}[t]
\includegraphics[]{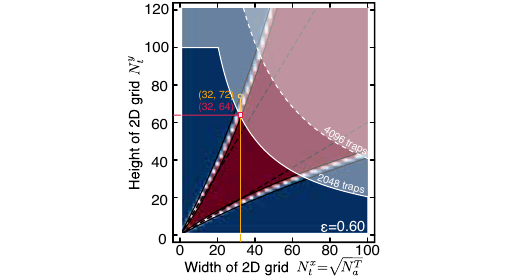}
\caption{
\label{fig:operational_performance}
\textbf{Mean success probability for 2D grids.}
Mean probability of successfully preparing a configuration of $N_a^T=\sqrt{N_a^T}\times\sqrt{N_a^T}$ atoms in a grid of $N_{t}^x\times N_{t}^y$ traps in the presence of loss. The loading efficiency is set to $\epsilon=0.60$. 
The midline curve of the transition region scales quadratically with the width $N_t^x=\sqrt{N_a^T}$ of the grid, $N_t^y\sim\mathcal{O}((N_t^x)^2)=\mathcal{O}(N_a^T)$, and thus linearly with the number of traps. In the presence of loss, the number of traps needed thus scales as the 3/2 power of the number of desired atoms, $N_t\sim\mathcal{O}((N_a^T)^{3/2})$.
}
\end{figure}

In the presence of loss, the near-success region is achieved for a larger numbers of traps~(Fig.~\ref{fig:operational_performance}) than in the lossless case. These additional traps are required to load the surplus atoms needed to compensate for loss during control operations. The transition curve evaluated at $\bar{p}\approx0.5$ shows that the height of the trap array scales approximately quadratically with the width of the trap array, $N_t^y\sim \mathcal{O}((N_t^x)^2)=\mathcal{O}(N_a^T)$, and thus linearly with the number of atoms. Choosing $N_t^y\approx\eta_0 + \eta_1 N_a^T$, we find that $\eta_0 = 0.01564$ and $\eta_1 = 1.579$ from linear regression.
This relationship implies that the number of traps required to assemble a configuration of $N_a^T$ atoms scales as $N_{t}=N_{t}^x\times N_{t}^y\sim \mathcal{O}((N_a^T)^{3/2})$, in contrast to the linear scaling observed in the absence of loss. This result is consistent with the red-rec algorithm solving approximately $N_t^x=\sqrt{N_a^T}$ columns, each column being solved with the 1D exact algorithm, which scales as $N_t^y\sim \mathcal{O}(\sqrt{N_a^T})^2$~(see Fig.~\ref{fig:1d_loss_a}).

\begin{figure}[t!]
\includegraphics[]{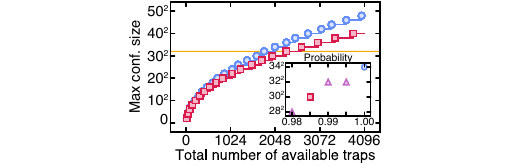}
\caption{
\label{fig:max_config_size}
\textbf{Largest achievable configuration of atoms.}
Maximum configuration size achievable with near-certain success in the absence (blue disks) and presence (red squares) of loss for a loading efficiency of $\epsilon=0.6$. Preparing a configuration of $32^2=1{,}024$ atoms with near-certain success requires $2{,}304$ traps. Inset: The maximum configuration size achievable with $2{,}048$ traps 
increases with the survival probability during displacement and transfer operations. We choose $p_\nu=p_\alpha$.
The red square is the data point associated with the specific survival probability used in the performance study. The purple triangles are the data points associated with other typical values for the survival probability.
}
\end{figure}

The largest configuration of atoms that can be prepared with near-certain success increases with the number of traps~(Fig.~\ref{fig:max_config_size}) and the survival probability during control operations. Fixing the total number of traps to be $N_t^{max}=2{,}048$~($32\times64$), the mean success probability for preparing a configuration of $32\times32$ atoms is $\bar{p}=0.21(1)$. Increasing the total number of traps to $N_t=2{,}304$~($32\times72$) increases the mean success probability to $\bar{p}=0.993(2)$. 
Increasing the survival probability during control operations from $p_\nu=p_\alpha=0.985$ to $p_\nu=p_\alpha=0.990$ increases the mean success probability from $\bar{p}=0.21(1)$ to $\bar{p}=0.996(4)$~(inset of Fig.~\ref{fig:max_config_size}). Further increasing the total number of traps to be $N_t^{max}=4{,}096$ limits the configuration size for near-certain success to $N_a^T=40^2=1{,}600$ atoms. The results highlight once again the challenges of scaling to large configuration sizes.

\begin{figure}[t]
\includegraphics[]{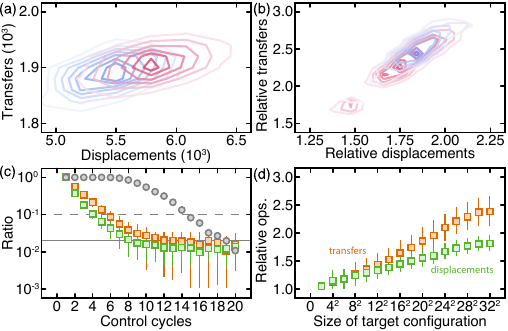}
\caption{
\label{fig:lossy}
\textbf{Control operations for 2D grids.}
(a)~Distribution of the total number of transfer and displacement operations during the first reconfiguration cycle of successful (blue) and unsuccessful (red) protocols. The protocols seek to prepare a configuration of $N_a^T=32\times32$ atoms in an array of $N_t=32\times64$ traps. During the first cycle, unsuccessful protocols perform more operations than successful protocols. 
(b)~Distribution of total transfer and displacement operations performed during all the subsequent reconfiguration cycles expressed relative to the first reconfiguration cycle. Successful protocols perform relatively more displacement and transfer operations than unsuccessful protocols during their subsequent reconfiguration cycles.
(c)~Probability that a successful protocol will perform at least $N_c$ control cycles (shaded disks). Less than $10~\%$ ($2~\%$) of successful protocols perform more than $15~(19)$ reconfiguration cycles. The total numbers of transfer (orange) and displacement (green) operations performed during successful protocols, expressed relative to the first cycle, monotonically decreases. 
(d)~Given a fixed overhead factor, here $\eta=2$ with $N_t^y=\eta N_t^x$), these numbers increase with the desired configuration size: increasingly more operations are performed in subsequent reconfiguration cycles to compensate for loss.
}
\end{figure}

To identify the conditions required for red-rec to succeed, we compute the mean total number of displacement and transfer operations performed during each reconfiguration cycle~(Fig.~\ref{fig:lossy}). We restrict our analysis to the problem of preparing a 2D configuration of $N_a^T=32\times 32$ atoms in a static array of $N_{t}=32\times64$ traps.

Firstly, we observe that successful protocols perform fewer transfer and displacement operations during their first reconfiguration cycle~(Fig.~\ref{fig:lossy}a) and more control operations in all their subsequent reconfiguration cycles than unsuccessful protocols~(Fig.~\ref{fig:lossy}b). 
Secondly, we observe that the relative total number of control operations decreases in later reconfiguration cycles. That is, fewer operations lead to less loss, and thus fewer atoms to be replaced, \emph{ad infinitum}, so that the late cycles perform fewer control operations to replace fewer lost atoms~(Fig.~\ref{fig:lossy}c).
Thirdly, we observe that the relative numbers of displacement and transfer operations increase with the size of the configuration of atoms~(Fig.~\ref{fig:lossy}d). 
We explain the increase in the relative numbers of displacement operations by the fact that, in larger arrays, surplus atoms left in their original positions must travel large distances to replace lost atoms in subsequent cycles. Because more displacement operations result in greater loss, more reconfiguration cycles are required to replace the lost atoms. We explain the increase in the relative numbers of transfer operations by the fact that receiver columns in late reconfiguration cycles are typically paired with multiple donor columns. Each pairing requires executing multiple redistribution sequences, each of which requires extracting and implanting the atoms.

\section{Reducing mean wait time via configuration rejection}~\label{sec:thresholding}

Besides improving the mean success probability of preparing a target configuration of atoms, an important operational requirement is increasing the duty cycle of experiments by reducing the mean wait time between two successful reconfiguration protocols. 
The mean success probability depends on the number of atoms in the initial configuration. These atoms are needed to prepare the target configuration of atoms and replace atoms lost during the multiple reconfiguration cycles. 

\begin{figure}[t]
\includegraphics{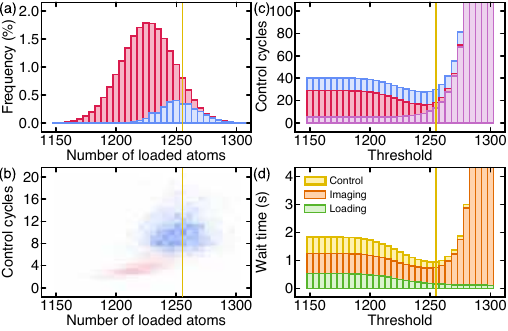}
\caption{
\label{fig:thresholding}
\textbf{Reducing mean wait time via configuration rejection.}
(a)~Stacked distributions of the number of atoms loaded into a square-lattice array of $N_t=32\times64=2{,}048$ traps for successful (blue) and unsuccessful (red) protocols when preparing a configuration of $N_a^T=32\times 32=1{,}024$ atoms. The threshold (vertical orange line) indicates the number of atoms, $N_a^{thresh}$, below which configurations are rejected.
(b)~Distribution of the number of control cycles per control protocol and the number of atoms loaded for successful (blue) and unsuccessful (red) protocols. Integrating over control cycles returns the stacked distributions of (a).
(c)~Number of measurements performed between two successful protocols for various threshold values. Measurements are performed whenever the algorithm fails to load more than $N_a^{thresh}$ atoms (purple), or fails (red) or succeeds (blue) in preparing the target configuration given more than $N_a^{thresh}$ atoms.
(d)~Stacked wait times between two successful protocols for various threshold values. The time is either spent loading the magneto-optical trap (green), imaging configurations of atoms (orange), or performing control operations (yellow). The threshold (vertical orange line) is chosen so as to minimize the total wait time.
}
\end{figure}

We partition the distribution of the number of atoms in the initial configuration partitioned for successful and unsuccessful protocols~(Fig.~\ref{fig:thresholding}a). We identify a first threshold below which a protocol is guaranteed near-certain failure ($\bar{p}\leq0.02$) and a second threshold below which a protocol is more likely to fail than to succeed. 
Rejecting configurations containing fewer atoms than either one of those two thresholds would increase the probability of successfully solving the reconfiguration problems for the retained configuration. However, even though the probability of success increases with the number of atoms in the initial configuration, the probability of measuring configurations with such a large number of atoms decreases. Thus, rejecting these configurations monotonically decreases the probability of success. Still, because unsuccessfully solving a problem might require multiple reconfiguration cycles~(Fig.~\ref{fig:thresholding}b), and thus more time than sampling a new configuration, there exists an optimal rejection threshold (orange vertical line in Fig.~\ref{fig:thresholding}) that minimizes the mean wait time between two successful protocols. This threshold is obtained by discarding as many unsuccessful configurations as possible, while rejecting as few successful configurations as possible without letting the number of measurements diverge~(Fig.~\ref{fig:thresholding}c). 

The mean number of measurements needed to sample a configuration of atoms that contains more atoms than the threshold is $1/P(N_a^0 \ge N_a^{thresh})$, where $P(N_a^0 \ge N_a^{thresh})$ is given by the cumulative distribution function of the binomial distribution. The mean amount of time spent preparing a configuration of atoms containing more atoms than the threshold is therefore $t_{MOT} + t_{image}/P(N_a^0 \ge N_a^{thresh})$. Here,  $t_{MOT}$ is the time needed to cool and trap atoms in a magneto-optical trap~(MOT), and $t_{image}$ is the time needed to image the configuration of atoms. We calculate the time elapsed between successful preparations of the target configuration by considering the MOT loading time, the time required to measure a configuration of atoms (imaging time), and the time required to execute the control operations. We take the MOT loading time to be $t_{MOT}=100~\text{ms}$, the time required to image the static array to be $t_{image}=20~\text{ms}$, and the execution time of parallel transfer and displacement operations to be $t_\alpha=15~\mu\text{s}$ and $t_\nu=67~\mu \text{s}$, respectively~(see Sec.~\ref{sec:parameters}). Atoms are loaded into the MOT and then continuously imaged in the static trap array until a configuration with sufficient atoms is measured. The MOT loading time is thus incurred only once for each attempted reconfiguration protocol.

From our numerical analysis, we find that the threshold value at which the minimum wait time occurs is $N_a^{thresh}=1{,}255$ atoms. As a result, a total of $87.6~\%$ of all configurations are discarded. Among these discarded configurations, $90.7~\%$ of them would have otherwise resulted in failure. After thresholding, $61.2~\%$ of control protocols result in success. For a configuration of $N_a^T=32\times 32$ atoms in a square-lattice array of size $N_{t} = 32\times 64$, thresholding reduces the mean time between successful preparations of the target configuration by a factor of approximately 2, from $1.84~\text{s}$ to $0.932~\text{s}$. We also observe that post-thresholding, the distributions in the mean number of reconfiguration cycles and the control time for successful and unsuccessful protocols are similar~(Fig.~\ref{fig:thresholding}d). These results indicate that achieving success depends mostly on random realizations of atom loss rather than on the number of atoms in the initial configuration. 

\section{Conclusion}\label{sec:conclusions}
We introduced the red-rec algorithm to solve atom reconfiguration problems on lattices, specifically focusing on grids. The red-rec algorithm reduces reconfiguration problems on grids to simpler reconfiguration problems on individual columns and pairs of columns. These problems are solved quickly and efficiently using simple algorithms. Individual columns are solved using the exact 1D algorithm; pairs of donor-receiver columns are solved heuristically by redistributing atoms from a donor column to a receiver column.

Although we describe the red-rec algorithm over grids, it can be extended to other lattice geometries, as well as other sub-lattice geometries embedded in the grid. Examples are checkerboard sub-lattices and oblique lattices with rhomboidal boundaries, including triangular lattices (with two generator vectors of equal norm oriented at a relative angle of $\pi/3$, e.g., see Ref.~\cite{Tian2023}). 

We numerically compared the performance of the red-rec algorithm against exact and approximation algorithms, both in the absence and presence of loss.
In the absence of loss, we showed that red-rec performs well at minimizing displacement operations. However, it performs more transfer operations than the value achieved by the 3-approx algorithm (see Fig.~\ref{fig:comparison_to_exact} in App.~\ref{sec:exact_benchmarking}). Further reduction in transfer operations might be achieved by allowing for the redistribution of atoms in the target region and solving for blocks of columns containing more than two columns, e.g., containing two donors and one receiver.

Improved reconfiguration algorithms enable preparing large configurations of atoms for a given mean success probability. In the presence of loss, red-rec prepares configurations of $256$ atoms in $512$ traps with a mean success probability of $\bar{p}=0.913(2)$. Moreover, it prepares configurations of $1024$ atoms in $2048$ traps with a mean success probability of $\bar{p}=0.21(1)$. 
 
Equivalently, improved algorithms facilitate achieving fast preparation times for a given configuration sizes. Rejecting configurations of atoms containing fewer atoms than a given threshold enables even faster preparation times. 

The red-rec algorithm is readily and efficiently implementable. It is also compatible with standard acquisition and control systems. Future work will focus on extending red-rec to other geometries and higher dimensions, implementing it on parallel-processing computing devices, and quantifying its operational performance experimentally.  

In addition, further opportunities exist to develop exact and approximation algorithms that simultaneously optimize multiple-objective functions. 
These algorithms might in turn guide the development of improved exact-heuristic algorithms for atom reconfiguration problems, which might also support other domains of applications.

Finally, we refer the interested reader to our complementary study on the assignment-rerouting-ordering (aro) algorithm~\cite{ElSabeh2023}. The aro algorithm exploits subroutines to improve the performance of typical assignment algorithms. Among others, we provide a constructive algorithm to find a partial ordering of displacement trajectories that guarantee that each atom is displaced at most once. 

The source code for the benchmarking module and reconfiguration algorithms will be made available upon reasonable request. 


\section{Acknowledgements}
We acknowledge early contributions from Hongru Xiang, Drimik Roy Chowdhury, Jin Tian (Benny) Huang, and Ho June (Joon) Kim. This work was supported by the Canada First Research Excellence Fund (CFREF).
Amer E. Mouawad's work was supported by the Alexander von Humboldt Foundation and partially supported by the PHC Cedre project 2022 ``PLR''. Research by Stephanie Maaz and Naomi Nishimura was supported by the Natural Sciences and Research Council of Canada.

\begin{appendix}

\section{Exact and approximation algorithms}\label{sec:exact_algorithms}

A reconfiguration algorithm provides a deterministic procedure that can be implemented as a sequence of programmable instructions to compute a valid solution to a given reconfiguration problem.
\emph{Exact algorithms} return solutions that are provably optimal, whereas \emph{approximation algorithms} return approximate solutions with provable guarantees on their distance to an optimal solution. In other words, $\mu$-approximation algorithms return solutions that are at most (at least) $\mu$ times the optimal value in the case of a minimization (maximization) problem. 

We have previously studied exact and approximation algorithms to solve reconfiguration problems on graphs from a theoretical standpoint~\cite{Cooper2024}. We now briefly describe four such algorithms: (1)~
the exact tree algorithm, (2)~the exact 1D algorithm, (3)~the minimum-weight maximum-cardinality matching (MWMCM) algorithm, and (4)~the 3-approximation (3-approx) algorithm. 

The \emph{exact tree algorithm} is an exact reconfiguration algorithm on trees, i.e., graphs in which any pair of vertices is connected by exactly one path, which is defined as a finite sequence of edges. The exact tree algorithm returns a sequence of elementary displacement operations whose number is the minimum required to solve the reconfiguration problem. These operations can be further batched into a sequence of parallel control operations that minimize the number of control steps or total reconfiguration time. This algorithm runs in time linear in the number of vertices in the tree~\cite{Calinescu2007}, and is readily amenable to efficient implementation on computing hardware.

The \emph{exact 1D algorithm} is an exact reconfiguration algorithm on linear chains. While it is possible to execute the exact tree algorithm on linear chains (given that linear chains are trees), the exact 1D algorithm is computationally more efficient, even if marginally. If the number of atoms is equal to the number of target traps ($\Delta N_a=N_a - N_a^T = 0$), then there exists a unique assignment of atoms to target traps because atoms cannot move past each other. The assignment can be computed in time linear in the number of traps in the chain. In this case, the exact 1D algorithm assigns the furthermost atoms to the furthermost target traps, from the edges to the center of the array, in order of their position. Such an assignment minimizes the total number of displacement operations in a single extraction-displacement-implantation (EDI) sequence~(Fig.~\ref{fig:reconfiguration_protocol}b). This EDI sequence simultaneously extracts, displaces, and implants the atoms that are not located in their target traps. In addition, it leaves the atoms that do not need to be displaced in their respective static traps. The exact 1D algorithm is also capable of solving reconfiguration problems where $\Delta N_a > 0$ in time linear in the number of traps in the chain~\cite{DBLP:journals/dm/KarpL75}.

\begin{figure}[t]
\includegraphics[]{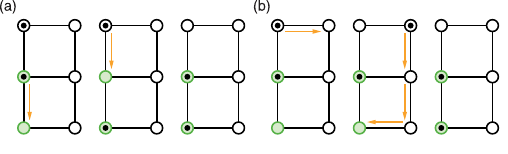}
\caption{
\label{fig:example}
\textbf{Minimal example for exact and approximation algorithms.}
(a)~The MWMCM algorithm minimizes the total displacement distance without concern for the number of atoms displaced, here displacing two atoms (black circles), each by an elementary step.
(b)~The 3-approx algorithm approximately (in this specific example, exactly) minimizes the number of atoms displaced without concern for the total displacement distance, here displacing one atom by four elementary steps.
}
\end{figure}

The \emph{minimum-weight maximum-cardinality matching algorithm} (MWMCM, see Fig.~\ref{fig:example}a for a minimal example) is an assignment algorithm that can be easily adapted into an exact reconfiguration algorithm valid on any graph. This algorithm returns an optimal solution to the problem of minimizing the total distance traveled by atoms, i.e., the total number of displacement operations, and it does so in polynomial time. The atoms in the initial and target configurations of atoms, associated with occupied traps in the static trap array, naturally define a weighted bipartite graph upon which this problem may be solved. Here, the weight of an edge represents the length of a shortest path in the original graph. This algorithm does not consider transfer operations, i.e., it does not distinguish between moving one atom twice or two atoms once. Indeed, for any pair of initial and final traps, the algorithm chooses any arbitrary path among the set of all of shortest paths, irrespective of the number of atoms on that path. The operational performance of the algorithm could be further improved by choosing the shortest paths that minimize the number of atoms displaced. This algorithm has a running time that is cubic in the number of vertices in the graph~\cite{kuhn1955hungarian}.

The \emph{3-approximation algorithm} (3-approx, see Fig.~\ref{fig:example}a for a minimal example), which is valid for any graph, returns a 3-approximate solution to the problem of minimizing the total number of atoms displaced. This problem is \textsf{NP}-complete even on grid graphs~\cite{Calinescu2007}. 
Given a graph, the algorithm partitions the graph into a collection of \textit{balanced trees}, where a balanced tree is a tree containing as many atoms as target traps. It does so by solving the U-Steiner problem~\cite{Calinescu2007}, removing edges in order to transform the graph into a collection of balanced trees. It then applies the exact tree algorithm to each tree. This algorithm could be further improved with regards to the total number of displacement operations by choosing a weighting function that penalizes convoluted paths through trees. In addition, the function could consider ``useful" trees to be those that reduce the number of displacement operations, or more generally, the total weighted distance traveled by all atoms. 

The MWMCM algorithm, an exact algorithm that minimizes total displacement operations, and the 3-approx algorithm, an approximation algorithm that seeks to minimize transfer operations, both target the minimization of a single-objective function. However, they are not optimal for any non-trivial linear combinations of both parameters, i.e., multiple-objective functions.
To the best of our knowledge, there exist no known optimal or approximation algorithms on grids, graphs, or any other relevant geometries that solve reconfiguration problems for multiple-objective functions. Examples of such functions might minimize a linear combination of the numbers of displacement and transfer operations, in polynomial time. In fact, we know that such algorithms do not exist unless $\textsf{P} = \textsf{NP}$, because minimizing the number of transfer operations alone is \textsf{NP}-complete~\cite{Calinescu2007}. Since optimizing for the multiple-objective function is a more general problem, solving it in polynomial time would imply a polynomial-time algorithm for minimizing the number of transfer operations. Our proposed red-rec algorithm heuristically seeks to simultaneously minimize the total number of displacement and transfer operations.

\section{Benchmarking performance against exact algorithms}\label{sec:exact_benchmarking}

\begin{figure}[t]
\includegraphics[]{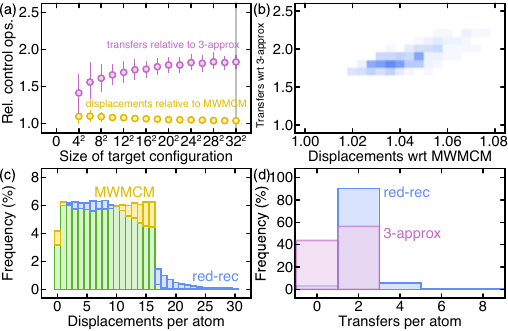}
\caption{
\label{fig:comparison_to_exact}
\textbf{Relative numbers of control operations in the absence of loss.}
(a)~Relative numbers of transfer and displacement operations performed by the red-rec algorithm expressed relative to the 3-approx and MWMCM algorithms, respectively. The numbers are reported for varying configuration sizes at constant overhead factor $\eta=2$ with no surplus atoms ($N_a^0 = N_a^T$). The relative number of transfer operations increases with configuration size.
(b)~Distribution of the number of transfer operations relative to 3-approx and displacement operations relative to MWMCM for preparing a configuration of $32\times32$ atoms in an array of $32\times 64$ traps. Red-rec extracts and implants significantly more atoms than the 3-approx. 
(c)~Distribution of the number of displacement operations performed per atom for the red-rec (blue) and MWMCM (yellow) algorithms. Red-rec displaces some atoms more times than the largest number of displacements achieved by MWMCM.
(d)~Distribution of the number of transfer operations per atom for the red-rec (blue) and 3-approx (purple) algorithms. As opposed to the 3-approx that extracts and implants each atom at most once, red-rec extracts and implants the same atom more than once.
}
\end{figure}

To obtain bounds on the minimum number of displacement and transfer operations required to solve a given atom reconfiguration problem, we further benchmark the red-rec algorithm against the MWMCM and 3-approx algorithms. MWMCM minimizes the total number of displacement operations, and 3-approx bounds the total number of transfer operations by at most three times its optimal value~(see App.~\ref{sec:exact_algorithms} for a description of these two algorithms). 

We specifically focus on the problem of preparing a square-compact configuration of $N_a^T=\sqrt{N_a^T}\times \sqrt{N_a^T}$ atoms in the center of rectangular array of $N_t=\sqrt{N_a^T} \times 2 \sqrt{N_a^T}$ traps, i.e., with twice as many traps as desired atoms ($\eta=2$). We set the number of surplus atoms to zero by choosing the number of atoms in the initial configuration to be equal to the number of atoms in the target configuration, i.e., $N_a^0=N_a^T$.

The red-rec algorithm performs slightly more displacement operations than the optimal value achieved by the MWMCM algorithm. The relative numbers of displacement operations decrease with increasing system size and reach a value of $1.04(1)$ for a configuration of $N_a^T=32^2=1{,}024$ atoms~(Fig.~\ref{fig:comparison_to_exact}a). 
The slight excess in displacement operations results from a small fraction of atoms, such as those redistributed across distant pairs of donor-receiver columns, undergoing up to twice as many displacements as the maximum per-atom displacement value  measured for MWMCM~(Fig.~\ref{fig:comparison_to_exact}b). 
The number of displacement operations is positively correlated with the number of transfer operations~(Fig.~\ref{fig:comparison_to_exact}c).
Compared to the 3-approx algorithm, the red-rec algorithm performs 1.5 to 2.0 times more transfer operations. We note that transfer operations include both extraction and implantation operations, so that the total number of transfer operations is even by construction, that is, all extracted atoms are implanted back in the static trap array.
The relative numbers of transfer operations increase with increasing system size, up to a value of $1.8(1)$ for a configuration of $N_a^T=32^2$ atoms~(Fig.~\ref{fig:comparison_to_exact}a). 
Whereas the 3-approx algorithm leaves nearly half of the atoms idle, the red-rec algorithm moves nearly all atoms at least once as it sequentially reconfigures each column at least once with fewer than 7\% of the columns getting reconfigured more than twice~(Fig.~\ref{fig:comparison_to_exact}d).

\section{Termination of the red-rec algorithm}\label{app:termination_proof}
We claim that, as long as the problem is solvable, there always exists at least one pair of donor-receiver columns for which at least one atom can be redistributed from the donor to the receiver (where at least one distribution row exists between the columns). This invariant guarantees that the red-rec algorithm indeed terminates. To see why, suppose instead that there are no distribution rows between any of the donor-receiver pairs. Consider the donor-receiver pair that minimizes the distance between the donor column and the receiver column and denote this pair by $(c_d, c_r)$. By assumption, these two columns cannot be adjacent as otherwise we have at least one distribution row. Assume that distribution rows between adjacent columns are not restricted to the storage region. Moreover, since we select $(c_d, c_r)$ by minimizing the distance between the two columns, there cannot exist any donor column or receiver column between $c_d$ and $c_r$. Otherwise, we can find a donor-receiver pair consisting of two columns separated by fewer columns than $(c_d, c_r)$, contradicting our choice of pair. Hence, all  columns between $c_d$ and $c_r$ must be neutral columns that have therefore been previously solved. This, in turn, implies that the storage region between $c_d$ and $c_r$ must be free of atoms and each row in this region can be used as a distribution row.

\end{appendix}


%

\end{document}